# Synthesis and characterization of gold-coated nanodiamonds through green chemistry as potential radiosensitizers for proton therapy

Edgar Mendes [1,9], Ana Belchior [1,2], Federico Picollo [3,4], Marta M. Alves [5,6], Rodica Mihaela Dinica [7], Maria Joao Moura [8], Teresa Pinheiro [2,9]* and Maria Paula Cabral Campello [1,2,*]

[1] Centro de Ciências e Tecnologias Nucleares, Instituto Superior Técnico, Universidade de Lisboa, Campus Tecnológico e Nuclear, Estrada Nacional 10, Km 139.7, 2695-066 Bobadela, Portugal; edgar.mendes@tecnico.ulisboa.pt , anabelchior@ctn.tecnico.ulisboa.pt, pcampelo@ctn.tecnico.ulisboa.pt

[2] Departamento de Engenharia e Ciências Nucleares (DECN), Instituto Superior Técnico, Universidade de Lisboa, Estrada Nacional 10, 2695-066 Bobadela, Portugal; edgar.mendes@tecnico.ulisboa.pt, anabelchior@ctn.tecnico.ulisboa.pt; teresa.pinheiro@tecnico.ulisboa.pt, pcampelo@ctn.tecnico.ulisboa.pt

[3] Physics Department, National Institute of Nuclear Physics, Section of Turin, University of Turin, Via P. Giuria 1, 10125 Turin, Italy; federico.picollo@unito.it

[4] Nanomaterials for Industry and Sustainability (NIS) Inter-Departmental Centre, University of Turin, Via Quarello 15/A, 10125 Turin, Italyfederico.picollo@unito.it

[5] Centro de Química Estrutural (CQE), Institute of Molecular Sciences (IMS), Instituto Superior Técnico, Universidade de Lisboa, Av. Rovisco Pais, 1049-001, Lisboa, Portugal; martamalves@tecnico.ulisboa.pt

[6] Departamento de Engenharia Química (DEQ), Instituto Superior Técnico, Universidade de Lisboa, Av. Rovisco Pais, 1049-001 Lisboa, Portugal; martamalves@tecnico.ulisboa.pt

[7] Faculty of Sciences and Environment, Department of Chemistry Physical and Environment, "Dunărea de Jos" University of Galati, 111 Domnească Street, 800201 Galati, Romania rodica.Dinicas@ugal.ro

[8] Instituto Nacional de Investigação Agrária e Veterinária, I.P. - Laboratório Químico Agrícola Rebelo da Silva Tapada da Ajuda, P-1301-596 Lisboa. Portugal; mjoao.mouras@iniav.pt

[9] iBB-Institute of Bioengineering and Biosciences, Instituto Superior Técnico, University of Lisbon, 1049-001 Lisbon, Portugal; teresa.pinheiro@tecnico.ulisboa.pt

* Correspondence: pcampelo@ctn.tecnico.ulisboa (MPCC); teresa.pinheiro@tecnico.ulisboa.pt (TP)

**Featured Application:** Authors are encouraged to provide a concise description of the specific application or a potential application of the work. This section is not mandatory.

## 1. Introduction

Today, cancer patients have numerous treatment options, with surgery, chemotherapy, and radiotherapy being the most common approaches. Surgery aims to completely remove the tumor but is typically feasible only in early stages or when the tumor is located in an operable area. It can also serve palliative purposes or involve partial tumor removal, often followed by chemotherapy or radiotherapy [1]. Chemotherapy uses cytotoxic agents to inhibit cell division and growth by causing genetic damage or blocking essential enzymes required for cell replication. However, its primary drawback is the lack of selectivity for cancer cells, leading to significant side effects and prompting efforts to develop more targeted treatments [2]. Radiotherapy, on the other hand, uses ionizing radiation—mainly X-rays—to destroy cancer cells or halt their proliferation. This radiation interacts directly with molecules, breaking bonds, or indirectly by generating Reactive Oxygen Species (ROS) that cause structural damage. These interactions often lead to DNA damage, including Single Strand Breaks (SSBs), Double Strand Breaks (DSBs), and base damage [3]. Despite the availability of these treatments, there is still no definitive cure for cancer. Classically, a combination of two or more strategies is employed to improve outcomes. Thus, ongoing research is critical to develop new and more effective therapies.

Proton therapy utilizes proton beams with a distinctive depth-dose profile that concentrates the radiation dose at the final microns of their path, forming a Bragg peak. This enables highly precise targeting of the radiation dose, minimizing harm to healthy tissues, unlike X-rays, which deliver their maximum dose to superficial tissues. However, proton therapy has a Relative Biologic Effectiveness (RBE) of 1.1, making it only 10% more effective than conventional radiation therapy [4]. Additionally, it requires costly equipment and specialized expertise. Despite these limitations, proton therapy is experiencing global growth, with over 100 operational facilities and over 50 under development [5]. Consequently, investigating new radiosensitizers — agents that enhance the biological impact of radiation - is vital to improving the efficacy of cancer therapy in general and proton therapy in particular.

Nanodiamonds (ND) are particularly attractive for biomedical applications due to their biocompatibility, ease of functionalization, and ability to penetrate biological barriers, offering potential solutions to the limitations of conventional approaches in cancer theranostics [6-8]. Their unique properties, such as the presence of Nitrogen Vacancy (NV) centers, further enhance their applicability in biomedical sciences. NV centers, characterized by a nitrogen atom substituting a carbon atom and a nearby lattice vacancy, exist in two forms: the neutral $NV^0$ and the negatively charged $NV^-$. Of particular importance is the $NV^-$ center, which, due to its spin ground state (S = 1), can be spin-polarized by optical pumping and controlled via electron paramagnetic resonance. When excited with radiation in the range of 465 to 565 nm, the $NV^-$ center emits light in the red spectrum, making NDs highly photostable with low photobleaching compared to common fluorescent dyes (Figure 1). This photoluminescent property, combined with the ability to induce additional NV centers, through high-energy particle irradiation followed by annealing at 600–800°C, positions NDs as a valuable tool for bioimaging [7-16].

Nanodiamonds have garnered significant attention in research not only for their biomedical applications but also for their potential as dose detectors and radiosensitizers. Upon proton irradiation, their unique structure undergoes measurable changes, enabling them to serve as sensitive tools for detecting and quantifying radiation doses. This property is particularly valuable in fields such as radiation therapy and materials science, where precise dose measurements are critical. Furthermore, nanodiamonds have been investigated for their ability to enhance radiosensitivity during proton irradiation, complementing the broader exploration of diamond-based detectors for dosimetric purposes in proton therapy [17-20].

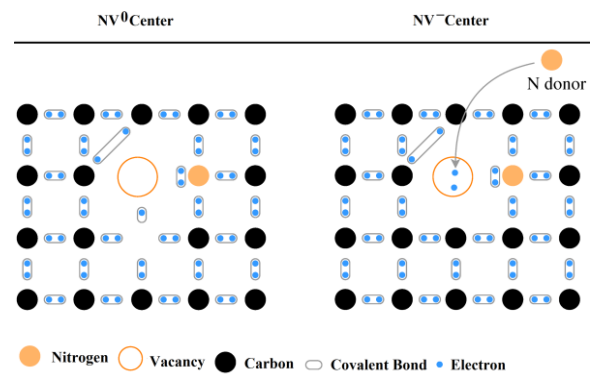

Figure1. $NV^0$ Center (Neutral): Formed by a nitrogen atom adjacent to a vacancy, with five electrons contributed by the center. These include three from the dangling carbon bonds around the vacancy (two of which form a bond and one remains unpaired) and two from the nitrogen lone pair. This state is non-magnetic. $NV^-$ Center (Negatively Charged): When an additional electron is donated (commonly from another nitrogen donor in the lattice), the NV center becomes negatively charged. This charge state exhibits unique magnetic properties. Adapted from [11]

Additionally, surface modifications enable the conjugation of therapeutic agents, such as drugs or nucleic acids, and imaging probes, such as fluorophores or radiolabels, turning nanodiamonds into multifunctional platforms for targeted drug delivery and real-time disease monitoring [6-8,21-25]. A particularly promising modification involves the conjugation of gold onto the surface of nanodiamonds. Gold nanoparticles are well known for having exceptional optical and electronic properties, which make them perfect for a variety of biomedical applications.

The resulting composite—nanodiamond-gold hybrids—represents a cutting-edge development in nanotechnology, offering a synergistic combination of the unique properties of nanodiamonds and gold nanoparticles. Nanodiamonds provide exceptional photostability, biocompatibility, and versatile surface chemistry, while gold nanoparticles contribute plasmonic and catalytic properties that enhance imaging, sensing, and therapeutic applications. This integration creates multifunctional systems with enhanced performance in bioimaging techniques, such as surface-enhanced Raman spectroscopy (SERS), photoacoustic imaging and multimodal imaging, leveraging the plasmonic properties of gold [26-29]. These hybrid nanoplatforms also demonstrate significant potential in biosensing and diagnostics, including immunosensors for the detection of cancer biomarkers like HER2, which exhibit high sensitivity and specificity in human serum samples [30]. Hybrid architectures, such as silica-encapsulated nanodiamonds coated with a gold shell, are effective in selective thermal ablation of cancer cells using near-infrared laser irradiation [31]. Additionally, their catalytic properties and DNA-based coupling mechanisms enable intracellular biosensing and therapeutic delivery, establishing nanodiamond-gold hybrids as versatile tools for advancing precision medicine [32].

Through a radiosensitization process, gold-conjugated nanodiamonds provide notable benefits in augmenting the effectiveness of radiation therapy, in addition to their contributions to imaging and photothermal therapy. Because of their high atomic number (Z), which increases the local deposition of radiation energy, gold nanoparticles are well-established radiosensitizers. Gold can promote the generation of secondary electrons when applied to the surface of nanodiamonds, which increases DNA damage in cancer cells when exposed to ionizing radiation [20, 33-37]. This effect is particularly advantageous for treating resistant tumors, where conventional radiation doses may be insufficient to achieve therapeutic outcomes. By leveraging gold's radiosensitizing properties, gold-functionalized nanodiamonds can potentially lower the required radiation dose, minimizing damage to surrounding healthy tissues while maintaining therapeutic efficacy. In proton therapy, AuNPs trigger mechanisms like the Coulomb nanoradiator effect and Auger cascades. Proton interactions with gold atoms strip electrons, generating reactive oxygen species (ROS) and X-rays, while secondary electron ejections via the Auger effect further increase tumor damage [38-41].

The physical, optical, and biological properties of nanoparticles strongly depend on their size, shape, surface functionalization, and the dielectric properties of their environment. These characteristics are intrinsically linked to the experimental conditions employed during their synthesis [42]

Herein, we developed hybrid gold-nanodiamond (NDAu) nanostructures using two types of commercial nanodiamonds (NDs) with median sizes of 50 nm and 230 nm. The NDs were annealed at 800°C for 2 hours in a nitrogen atmosphere, yielding annealed nanodiamonds (Ann NDs 50 and Ann NDs 230). These were subsequently oxidized in air at 500°C for 12 hours, producing oxidized nanodiamonds (Ox NDs 50 and Ox NDs 230) [20]. The gold coating on the NDs was synthesized using green chemistry methods with Nymphaea alba root extracts as reducing agents, in line with sustainable practices. Nymphaea alba, also known as the European white-water lily, is rich in phytochemicals whose composition varies by geographical location and plant part. These extracts are known for their antioxidant, anti-inflammatory, and hepatoprotective properties, as well as their traditional use in treating diabetes, urinary tract infections, and other ailments. The key bioactive effects are largely attributed to polyphenols such as ellagic acid and flavonoids like quercetin and kaempferol [43, 44].

Green chemistry emphasizes reducing the environmental and health impacts of nanoparticle production by avoiding toxic solvents and pollutants. It leverages biological molecules such as polyphenols, citric acid, flavonoids, and reductases found in plants, algae, or microorganisms [43-47]. These methods are versatile, cost-effective, and can incorporate biological waste from agricultural or food industries, lowering the carbon footprint of nanoparticle synthesis. Furthermore, green nanoparticles exhibit improved biocompatibility, as harmful chemical contaminants are avoided. Organic compounds from plant extracts integrated into the nanoparticle surfaces can enhance cellular uptake and confer additional beneficial properties, such as antimicrobial or anticancer activity [48].

While green chemistry has been widely used to produce metal or metal oxide nanoparticles, to our knowledge, this is the first report of hybrid gold-nanodiamonds synthesized using this approach. Comparable methods, such as the environmentally friendly synthesis of silver nanoparticles on nanodiamonds using tea polyphenols, have demonstrated the potential of such strategies [48].

In this study, the synthesized hybrid gold-nanodiamond underwent comprehensive physicochemical characterization. Their biological activity was evaluated through cytotoxicity assays, cell survivability and colony formation assessments, and cellular uptake studies. These evaluations were conducted using the A549 lung cancer cell line and the PANC-1 pancreatic cancer cell line, both critical cancers located near vital structures, complicating treatment due to the risk of damaging healthy tissues. Moreover, these cancers often exhibit resistance to conventional therapies, underscoring the need for innovative and targeted treatment approaches [49,50].

**2. Materials and Methods**

2.1. Reagents and materials

All chemicals and solvents were of reagent grade and were used without additional purification, unless stated otherwise. The Milli-Q water was produced from a Millipore system Milli-Q ≥18 MΩcm (with a Millipak membrane filter 0.22 μm). Tetrachloroauric (III) acid trihydrate 99% ($HAuCl_4 \cdot 3H_2O$) and sodium hydroxide (NaOH) were commercially acquired from Aldrich Chemical. The roots of the *Danube Delta Nymphaea alba* species were obtained from the Biosphere Reserve of Romania. and Environment, Department of Chemistry Physical and Environment, "Dunărea de Jos" University of Galati [45].

The A549 lung carcinoma epithelial cell line, the PANC-1 human pancreatic cancer cell line were purchased from the American Type Culture Collection. The cells were kept at 37 °C and 5% CO2 in a humidified atmosphere (Heraeus, Germany). All products for cellular studies, including the cellular media and supplements, were acquired from Gibco (Thermo Fisher Scientific), unless otherwise stated.

2.2. Synthesis of the Gold coated Nanodiamonds (NDAu)

2.2.1 Thermal treatment of the Nanodiamonds (ND)

Two commercially available types of nanodiamonds, with sizes of 50 nm (MSY 0-0.1) and 230 nm (MSY 0-0.5), were subjected to thermal treatments. Both were annealed at 800°C for 2 hours in a $N_2$ atmosphere, resulting in samples labeled Ann 50 nm ND and Ann 230 nm ND. Subsequently, a portion of the annealed nanodiamonds underwent oxidation at 500°C for 12 hours in air, forming samples Ox 50 nm ND and Ox 230 nm ND, respectively. These treatments were performed at the University of Turin by Federico Picollo [20].

2.3. Processing of the root extract used for the synthesis of gold-coating

Root extracts from *Nymphaea alba* were utilized to synthesize the gold coating on nanodiamonds[42,45]. A stock solution of the root extracts was prepared to be used in the gold coating synthesis process of the NDs. 2g of powder root extracts were dispersed in 120 mL of Milli-Q water and stirred during 1h at 30°C. The obtained extract suspension was cooled to room temperature and twice filtered through a Whatman No. 1 filter paper. The filtered yellow extract was used immediately to prevent any changes in its composition or loss of reactivity, which is crucial for the gold coating process on the nanodiamonds.

2.4. General procedure for the gold-coating

The NDs were dispersed in 25 mL of separate root extract solutions and subjected to ultrasonic irradiation for 10 minutes at room temperature, using an ultrasonic batch (FisherbrandTM S-Series FB15051) operating at 38 kHz and 100 W. This was followed by magnetic stirring for an additional 40 minutes. After stirring, a 25 mM $HAuCl_4$ water solution was introduced, causing the reaction mixture to immediately change color from bright yellow to brown. Then, a 0.1 M NaOH solution was added dropwise until the pH reached 8. At this point, a second volume of 25 mM $HAuCl_4$ water solution was gradually added while stirring at room temperature, leading to a color change from brown to gray-blue. The samples were incubated overnight at room temperature with continuous magnetic stirring.

After incubation, the solutions were washed twice with Mili-Q water at 6000 RPM for 20 minutes and then freeze-dried for 48 hours. Table 1 lists the acronyms for each gold-coated ND and the specific quantities of reagents and precursors used at each synthesis step.

Table 1: Type of modified NDs, ND precursor and volume of gold added

| Sample | Precursor | Volume of $HAuCl_4$ (mL) |
|---|---|---|
| Ann NDAu 50 | 28 mg Ann 50 nm ND | 2 + 8 |
| Ann NDAu 230 | 18 mg Ann 230 nm ND | 1.5+4.5 |
| Ox NDAu 50 | 27 mg Ox 50 nm ND | 2 + 8 |
| Ox NDAu 230 | 17.1 mg Ox 230 nm ND | 1.5 + 4.5 |

2.5. NDAu characterization

NDAu characterization involved Ultraviolet-visible spectroscopy (UV-Vis), Powder X-ray Diffraction (PXRD), Dynamic Light Scattering (DLS) and Zeta Potential ($\zeta$-potential), Attenuated Total Reflectance- Fourier Transform Infrared (ATR-FTIR), Particle-Induced X-ray emission (PIXE), Raman spectroscopy, Scanning Electron Microscopy (SEM) and Transmission Electron Microscopy (TEM). UVVis spectra were recorded in water from 200-900 nm using a Varian Cary 50 UV/Vis spectrophotometer at room temperature. PXRD analysis was performed with a Bruker D2 Phaser diffractometer, equipped with a Cu K$\alpha$ X-ray tube, in the 2θ range from 10° to 100°, with crystal size and parameters calculated using Scherrer's equation [51]. DLS and $\zeta$-potential measurements utilized a Malvern Zetasizer Nano ZS operating at a 173° angle with a 633 nm He-Ne laser. ATR-FTIR spectra were obtained using a Thermo Fisher Scientific FTIR Nicolet iS50 spectrometer with an ATR diamond crystal accessory. Elemental quantification of carbon was determined by dry combustion on an Thermo Unicam Flashsmart® NCS analyser at National Institute for Agrarian and Veterinary Research and the gold elemental quantification was performed by PIXE analysis using a 2 MeV proton beam from the 2.5 MeV Van de Graaff accelerator at CTN/IST, following acid digestion (1:3 nitric to hydrochloric acids molar ratios) of NDAu samples. Raman spectra were acquired with a Horiba LabRAM HR800 Evolution using a 532 nm solid-state laser at 20 mW output power. SEM analysis was conducted using a Phenom ProX G6 or Hitachi S-2400, operating with an electron beam at 15 or 20 kV, respectively, while TEM images were obtained using a JEOL 1200EX in collaboration with CiiEM at Egas Moniz School of Health and Science.

2.6. Cell studies

2.6.1. Cytotoxicity Studies and Cell Survival

The cytotoxicity of NDAus was evaluated using the colorimetric assay 3-[4,5-dimethylthiazol-2-yl]-2,5 diphenyl tetrazolium bromide-MTT assay-, at concentrations of 5, 10, 20, 50, 100, or 200 μg/mL in the A549 and PANC-1 cell lines, with absorbance measured at 570 nm using a Varioskan LUX scanning multimode reader (ThermoFisher Scientific, US). To address the intrinsic absorbance of NDAus, a luminescence-based ATP assay kit (Sigma-Aldrich) was also used to evaluate cytotoxicity at the same range of concentrations in both cell lines. The clonogenic assay was used for cell survival analysis. The cells, 100 cells per well, were seeded in 6-well plates and incubated overnight for adhesion. Then, the cells were incubated for 24 hours, with 20 or 50 μg/mL NDAus. Afterwards, the medium was replaced by fresh medium, and the cells return to the incubator for 10 days. Every 3 days the medium was replaced.

2.6.2 Cellular Uptake Studies

To evaluate NDAu cellular uptake, A549 cells were seeded on 100 nm thick silicon nitride membranes, allowed to adhere overnight, and incubated for 24 hours in medium containing either Ann NDAu 50 or Ox NDAu 50 at a concentration of 20 μg/mL. After incubation, cells were washed with cold PBS and freeze-dried. Gold distribution within the cells was analyzed using PIXE and EBS techniques at the nuclear microscopy facility of the Van de Graaff accelerator (CTN/IST). Data acquisition, spectrum processing, and concentration calculations were performed with the OMDAQ-3 program, with elemental concentrations reported as μg/g dry weight [52].

3. Results and Discussion

3.1. Surface modification of Nanodiamonds

Commercial nanodiamonds of two sizes (50 nm and 230 nm) were evaluated, each subjected to distinct thermal treatments. One set was annealed at 800°C for 2 hours in a nitrogen ($N_2$) atmosphere, producing annealed nanodiamonds (Ann ND 50 and Ann ND 230). A portion of these was further oxidized in air at 500°C for 12 hours, generating oxidized nanodiamonds (Ox ND 50 and Ox ND 230). This study explored the potential of *Nymphaea alba* root extracts to synthesize and stabilize AuNPs directly on the ND surface, leveraging the extracts as a green reducing agent for the *in situ* reduction of $HAuCl_4$ and stabilization of the nanoparticles, resulting in a well-distributed NDAu hybrid nanoplatform. The Ox ND exhibited significantly improved dispersibility in water, likely due to the higher presence of oxygen-containing functional groups on its surface, which enhanced interactions with the reaction medium and contributed to the greater mass of the Ox NDAu nanoplatforms. In contrast, the Ann NDAu nanoplatforms showed a lower mass compared to the Ox NDAu counterparts, particularly for the larger 230 nm particles, despite equivalent amounts of gold precursor and root extract being used (see Table S1).

Integrating UV-vis, infrared, and Raman spectroscopies with Powder X-ray Diffraction (PXRD), Proton Induced X-ray Emission (PIXE), Scanning Electron Microscopy (SEM), Transmission Electron Microscopy (TEM), and measurements of DLS and zeta potential revealed changes in surface chemical compositions and confirmed the efficiency of the gold coating process on various types of nanodiamonds, achieved through a green chemistry approach using *Nymphaea alba* root extracts

It is anticipated that NDs' poorly absorbing particles will have UV-vis spectra that are strongly influenced by $\lambda^4$, resulting in a curve that is more akin to exponential decay, with absorbance rising quickly as wavelength drops, especially in the UV region [53]. The noteworthy change in the UV-vis spectra of the gold-modified NDAus confirms the modifications made to the nanoparticles (see Figure S1). LSPR is the term for the oscillatory effect of the electrons on the metal surface caused by the intense interaction of gold nanoparticles with light at specific wavelengths [54]. Thus, the absorption observed at 545 nm in the spectra of the Ann NDAu 50 and Ox NDAu 50 is attributed to the formation of gold nanoparticles. The absorption peak in the case of Ann NDAu 230 and Ox NDAu 230 appears around 550 nm, suggesting the formation of larger gold nanoparticles, since the red-shift on the absorption maxima of the SPR bands of gold nanoparticles reflects the tendency of the formation of larger nanoparticles and/or agglomerates. Additionally, the impact of the plant extracts on all of the NDAu spectra is demonstrated by two absorption bands at <300 nm, which correspond to the polyphenolic/flavonoids molecules of the root extracts. This suggests that the gold nanoparticle present in the NDAus nanoplatforms is effectively reduced and stabilized by the polyphenols in the root extract [43]. The FTIR and UV-Vis spectra for pristine Ann ND 50 nanodiamonds, modified Ann NDAu 50 nanodiamonds, and extracts are shown in Figure 3.1 while spectra for the remaining samples (Ann ND 50, Ann ND 230, Ox ND 230, and their gold coatings) are in Figures S2-S4. The pristine samples exhibit common stretching vibrations and bands specific to annealed or oxidized NDs. Key absorption bands include a broad peak at 3420 cm$^{-1}$ (O-H stretch of water), 1618 cm$^{-1}$ and 1630 cm$^{-1}$ (O-H bending), 1384 cm$^{-1}$ (C-N stretching), and 1090 cm$^{-1}$ and 110 cm$^{-1}$ (C-O stretching) [55,56]. The C-N band at 1384 cm$^{-1}$ is less prominent in the Ox NDs, compared to the annealed counterpart, indicating effective nitrogen removal. The Ann ND shows smaller bands at 2920 cm$^{-1}$ and 2849 cm$^{-1}$ (CH$_2$ stretching), while the Ox ND has unique bands at 1774 cm$^{-1}$ (C=O stretching from ketones) and 1263 cm$^{-1}$ (C-O bending from esters). The NDAu spectra reveal peaks at 1700 cm$^{-1}$ (C=C stretching of carboxylic acids) and 1445 cm$^{-1}$ (C-O-H bending in flavonoids), which align with the presence of polyphenols and flavonoids, such as ellagic acid and quercetin, reported as key bioactive compounds in *N. alba*. Additional peaks at 1582 and 1508 cm$^{-1}$ (C=C in aromatic compounds) and 1366 cm$^{-1}$ (O-H bending of phenols) further suggest the successful interaction of these phytochemical components with the gold surface during the green synthesis process. This indicates that the *Nymphaea alba* extracts not only played a crucial role as reducing agents but also contributed functional chemical groups that may enhance the stability and bioactivity of the hybrid nanostructures. The presence of esters and polyols, evidenced by the peaks at 1111 and 1200 cm$^{-1}$, respectively, further supports the involvement of plant-derived molecules in surface functionalization, which may be essential for future biomedical applications [43,44,55,56].

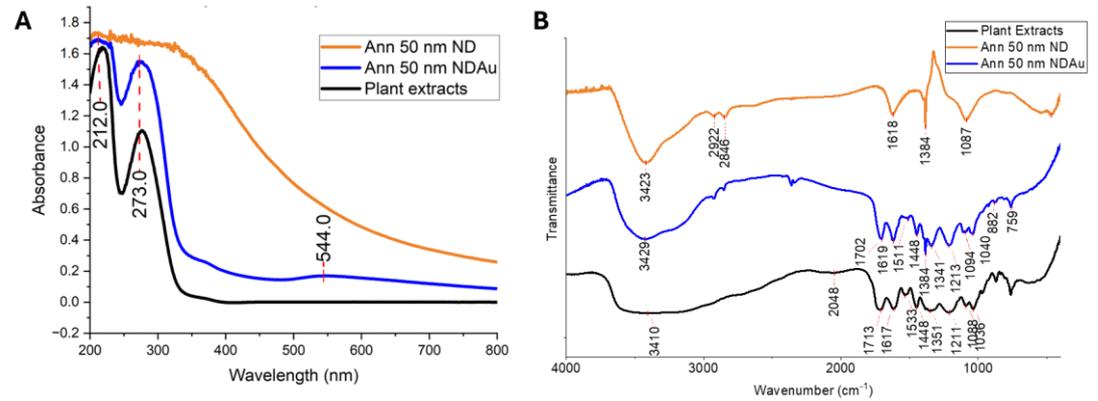

Figure 3.1: (A) UV-Vis and (B) FTIR spectra of 50 nm Ann ND, Ann NDAu and plant extracts

The Raman spectra for the 50 nm nanodiamonds and the gold-nanodiamonds are shown in Figure 3.2. Figures 3.2 A and 3.2B display the expected spectral bands: the diamond peak (red) at around 1320 cm$^{-1}$, which corresponds to sp$^3$ bonded carbon; the D band (green) at 1330-1340 cm$^{-1}$, indicating defects or disordered sp$^2$ carbon; and the G band (light blue) at around 1590 cm$^{-1}$, related to in-plane sp$^2$ carbon (graphene-like carbon). A fourth band, the D" band (dark blue), represents amorphous graphene oxide in the sample [57-59]. The fitting of bands for gold-coated nanodiamonds is not shown due to the loss of Raman signal, as gold does not produce a Raman signal and its presence reduces signal strength through attenuation and refraction. Notably, the diamond peak was significantly reduced or vanished, confirming the presence of a gold layer coating on the nanodiamonds. The Raman spectra for the 230 NDs and NDAus are shown in Figure 3.3. Unlike other NDAu samples, the Ann NDAu 230 sample exhibits a noticeable diamond peak (Figure 3.3 C), likely influenced by the gold shell, which alters Raman signals. Interestingly, this sample's significant diamond peak correlates with lower gold content on the surface, suggesting that the thickness of the gold shell critically affects Raman signal intensity in NDAu samples. Furthermore, Figure 3.3B highlights the effects of the oxidation process following the annealing of nanodiamonds. The D and G bands are considerably less intense than the diamond peak, visually confirming the selective etching of surface graphene-like carbon.

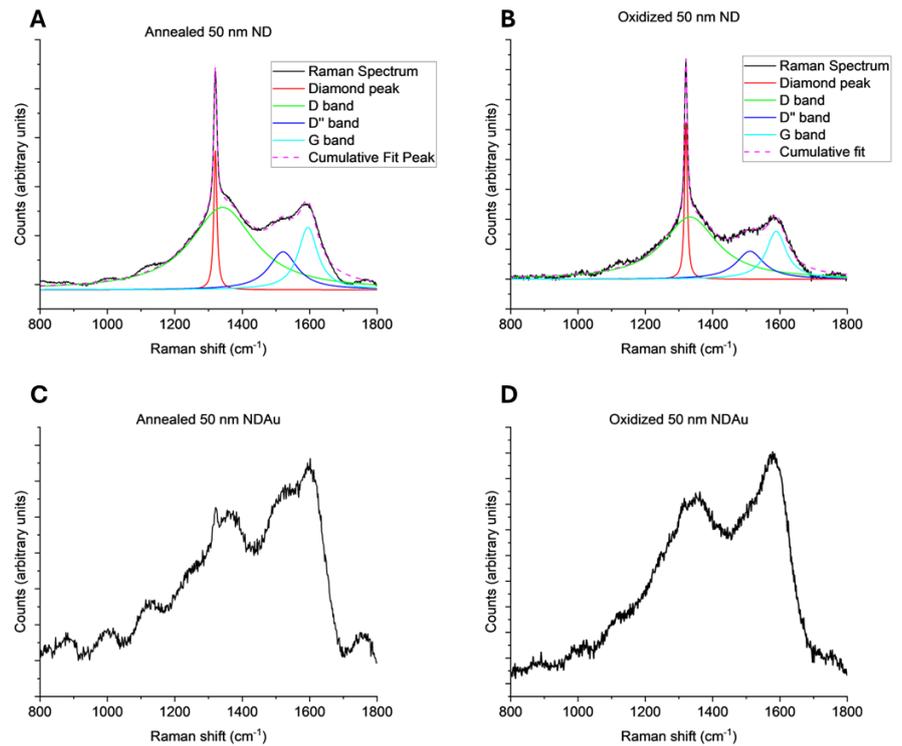

**Figure 3.2:** Raman spectra of 50 nm (A) Ann ND, (B) Ox ND, (C) Ann NDAu and (D) Ox NDAu. In unmodified nanodiamond spectra (A and B), 4 curves were deconvoluted from the spectra into the D peak (red), the D band (green), D'' band (dark blue) and G band (light blue).

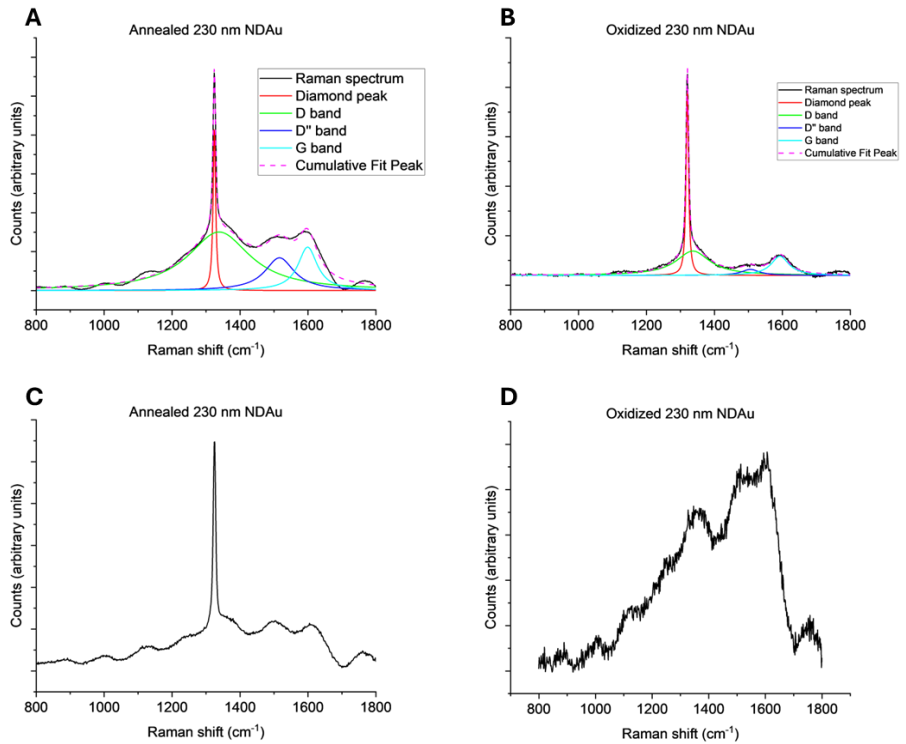

**Figure 3.3:** Raman spectra of 230 nm (A) Ann ND, (B) Ox ND, (C) Ann NDAu and (D) Ox NDAu. In unmodified nanodiamond spectra (A and B), 4 curves were deconvoluted

from the spectra into the D peak (red), the D band (green), D'' band (dark blue) and G band (light blue).

Table 2. presents the ratios derived from the deconvoluted Raman spectra of the 50nm and 230 nm nanodiamonds, providing insights into their carbon structure. These ratios, calculated from the maxima of each deconvoluted peak, reflect the contributions of different carbon species. Specifically, the ratios of the diamond peak to the G band and the diamond peak to the D band indicate the $sp^3/sp^2$ ratio, while the D-to-G ratio relates to the structural ordering of $sp^2$ carbon [60]. From the data, it is evident that the Ox NDs exhibit a higher relative diamond Raman signal compared to $sp^2$ phase carbon, as oxidation at lower temperatures enhances the diamond signal intensity. The D-to-G ratio, which increases as $sp^2$ carbon becomes more ordered, shows only a minimal rise at 500 °C, becoming more pronounced at higher temperatures [60,61]. Notably, the changes in the diamond peak ratios are more pronounced in the 230 nm NDs than in the 50 nm NDs. While the 50 nm samples exhibit approximately a 50% increase in the diamond peak ratio after treatment, the 230 nm samples demonstrate a more significant increase of around 160%, indicating a higher relative quantity of $sp^3$-bonded carbon compared to $sp^2$-bonded carbon in the larger samples. The greater size of the 230 nm NDs likely contributes to their higher diamond peak ratios in the annealed samples (Ann NDs), as the innermost regions of these larger particles were less affected by the annealing conditions, preserving more of the diamond core. In the Ox NDs, the D-to-G ratio decreases slightly as a result of the oxidation process. This decrease reflects a reduction in disordered $sp^2$ carbon, which is attributed to superficial etching that selectively enhances the $sp^3$-dominant diamond signal.

Table 2. Raman intesity ratios for the 50nm and 230nm NDS

| Sample | $\dfrac{\text{Diamond Peak}}{\text{G band}}$ | $\dfrac{\text{Diamond Peak}}{\text{D band}}$ | $\dfrac{\text{D band}}{\text{G band}}$ |
|---|---|---|---|
| Ann ND 50 | 1.70 | 2.24 | 1.32 |
| Ox ND 50 | 2.52 | 3.27 | 1.30 |
| Ann ND 230 | 2.79 | 3.76 | 1.35 |
| Ox ND 230 | 7.63 | 9.67 | 1.27 |

To assess the crystallinity and grain size of the nanodiamonds and their gold-coated counterparts, Powder X-ray Diffraction (PXRD) was performed on these samples within the 2θ range of 10° to 110° (Figure 3.4 and Figure….SI). The PXRD patterns of the nanodiamonds (NDs) exhibit characteristic diffraction peaks, which can be indexed to the diamond crystal planes (111), (220), and (311), at approximately 2θ = 44°, 2θ = 75°, and 2θ = 91°, respectively (Table 3). These values are consistent with the Crystallography Open Database (COD 9012304) and confirm the pure crystalline nature of annealed and oxidized NDs, with no other crystalline phases detected [62,63].

For the NDAus, the PXRD patterns reveal eight diffraction peaks. Three of these correspond to the crystal planes of diamond, while the remaining five are indexed to the (111), (200), (220), (311), and (222) lattice planes, characteristic of the face-centered cubic (FCC) structure of metallic gold, in agreement with the Crystallography Open Database (COD 9008463). The patterns confirm the purity of the samples, with no crystalline impurities detected.

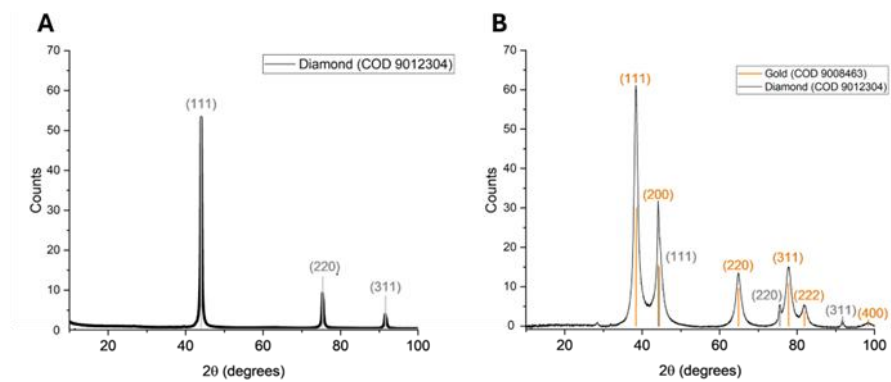

Figure 3.4: XRD pattern of 50 nm (A) Ann ND 50nm, (B) Ann NDAu 50nm

Using Scherrer's equation, the crystallite sizes were estimated based on the highest-intensity peaks. For the gold coating, the crystallite sizes were 7.31 nm and 7.76 nm for the Ann NDAu 50 and Ox NDAu 50 samples, respectively. For the diamond peaks, the crystallite sizes were 23.67 nm and 24.92 nm for the Ann ND 230 and Ox ND 230 nm samples, respectively. For the NDAu 230 samples, the gold crystallite sizes were calculated as 12.41 nm and 11.49 nm for the annealed and oxidized samples, respectively, while the diamond crystallite sizes were 33.85 nm and 36.81 nm, respectively. The gold crystallite sizes found here are in good agreement with values reported in the literature for other nanoparticles synthesized using green chemistry [64]. Further details regarding the diffraction peaks and crystallite size calculations for additional lattice planes are provided in Table S..

Table 3. PXRD Characterization of the 50nm and 230nm NDs and NDAus

| Sample | Phase | Lattice Parameter(Å)* | 2θ (°) | Crystallite Size (nm) |
|---|---|---|---|---|
| Ann ND 50 | Diamond | 3.56 | 44.04 | 23.67 |
| Ox ND 50 | Diamond | 3.56 | 44.04 | 24.92 |
| Ann ND 230 | Diamond | 3.56 | 44.05 | 33.85 |
| Ox ND 230 | Diamond | 3.56 | 44.04 | 36.81 |
| Ann NDAu 50 | Metallic Au | 4.07 | 38.43 | 7.31 |
| Ox NDAu 50 | Metallic Au | 4.07 | 38.43 | 7.76 |
| Ann NDAu 230 | Metallic Au | 4.07 | 38.34 | 12.41 |
| Ox NDAu 230 | Metallic Au | 4.07 | 38.32 | 11.49 |

Lattice parameter of the Crystallography Open Database for for diamond (COD 9012304), a = 3.5667 A°; and for Au (COD 9008463), a = 4.07825 A°

PIXE analysis revealed variations in the composition of modified nanodiamonds between annealed and oxidized samples. Table 4 presents the mass percentages of gold in NDAu, as determined through this technique. Trace amounts of other elements, such as calcium, iron, and potassium, were also detected (Table S...), but their contributions were minimal. The PIXE technique relies on the emission of characteristic X-rays caused by the relaxation of excited electron clouds, stimulated by energetic proton beams delivered by a particle accelerator. The low error values for all measurements (<1 %) underscore the high precision of the technique.

Regarding the NDAu 50 samples, the gold content was slightly higher in the oxidized samples (39.50 %) compared to the annealed ones (37.34 %), though the values were relatively close. This minor discrepancy may be attributed to the Ox NDs exhibiting a higher density of oxygen-containing surface groups. These functional groups likely enhanced their interaction with the reaction bath, resulting in stronger adsorption of gold ions and improved deposition. When it comes to the NDAu 230 samples, the differences between treatments were much more pronounced. The gold content in the Ox NDAu 230 (41.06 %) was more than double that of the Ann NDau 230 (19.62 %). This stark contrast highlights the critical role of oxidation in preparing larger nanodiamond surfaces for efficient gold deposition. The lower gold content in the annealed samples may reflect incomplete surface coverage, limited gold-nanodiamond interaction due to residual contaminants, or insufficient surface preparation. Conversely, the oxidative treatment likely removed impurities and activated the surface, increasing its reactivity and facilitating more uniform gold coverage.

**Table 4.** Carbon and Gold Content in the annealed (Ann) and oxidized (Ox) Nanodiamonds (ND)

| | Nanodiamond | Mean (w/w %) |
|---|---|---|
| Mean C[a] | **50 nm Ann ND** | 95.42 |
| | **50 nm Ox ND** | 45.28 |
| | **230 nm Ann ND** | 99.03 |
| | **230 nm Ox ND** | 77.64 |
| Mean Au[b] | **50 nm Ann NDAu** | 37.34 ± 0.06 |
| | **50 nm Ox NDAu** | 39.50 ± 0.04 |

| | |
|---|---|
| **230 nm Ann NDAu** | 19.62 ± 0.13 |
| **230 nm Ox NDAu** | 41.06± 0.06 |

[a] Carbon determined by dry combustion. [b] Gold deyermined by PIXE

These trends underscore the influence of particle size and surface-to-volume ratio on gold deposition. Smaller particles (50 nm), with their higher surface-to-volume ratios, appear to support more uniform gold coverage under both annealing and oxidation conditions. Larger particles (230 nm), however, seem to require oxidative treatment to expose sufficient reactive surface for comparable gold deposition. These results highlight the importance of surface chemistry and treatment methods in tailoring gold content and coating quality in nanodiamond modifications, particularly for applications that require precise control over coating properties, such as drug delivery, radiosensitization, or catalysis.

To further investigate these trends, SEM and TEM analyses were conducted to characterize the morphology of green-synthesized hybrid gold-nanodiamonds, focusing on their shape and size distribution. Figure 3.5 presents representative SEM (3.5 A-D) and TEM (3.5 E-H) images of the NDAu samples. The images reveal that NDAu tend to form small aggregates, which are more evident in the annealed samples. For the 50 nm annealed NDAu, the average aggregate size was approximately 28 nm, while the 230 nm annealed NDAu formed larger aggregates of around 98 nm. The oxidized 230 nm NDAu showed slightly smaller aggregates, averaging 78 nm contrasting with the oxidized 50 nm NDAu samples which showed a better dispersion and an average of yyyy (See Supplemetary Material Figure Sxxxx?).

The direct synthesis of gold nanoparticles (AuNPs) in a nanodiamond dispersion typically produces spherical or quasi-spherical particles. SEM and TEM images indicate that numerous small gold nanoparticles are distributed across the ND surfaces. However, the extent of AuNPs coverage varied, and not all diamond surfaces were fully coated. The gold nanoparticles exhibited a broad size distribution, suggesting uncontrolled growth after nucleation on the ND surface. This lack of size regulation can be attributed to the absence of capping agents and the intrinsic non-uniformity of the ND surface, which likely influenced nucleation and growth dynamics [65]. In future studies, it would be beneficial to explore variations in both the concentration and type of plant extracts used in the synthesis of AuNPs. Adjusting these parameters can significantly influence the size, shape, and functional properties of the resulting nanoparticles. For instance, a study by Teimuri-Mofrad et al. (2017) discusses the impact of different plant extracts on the green synthesis of AuNPs, highlighting how variations in extract composition can affect nanoparticle characteristics [66]. By systematically varying the extract type and concentration, we can optimize the synthesis process to tailor AuNPs for specific applications, such as drug delivery, catalysis, therapy or imaging [42,43, 66-68]

Energy dispersive X-ray spectroscopy (EDX) analysis confirmed that the NDAu were composed solely of carbon and gold, with no detectable impurities or artifacts. Red crosses in the Figures 3.5 A-D indicate the locations where EDS spectra were acquired. (see SI, Fig and table)

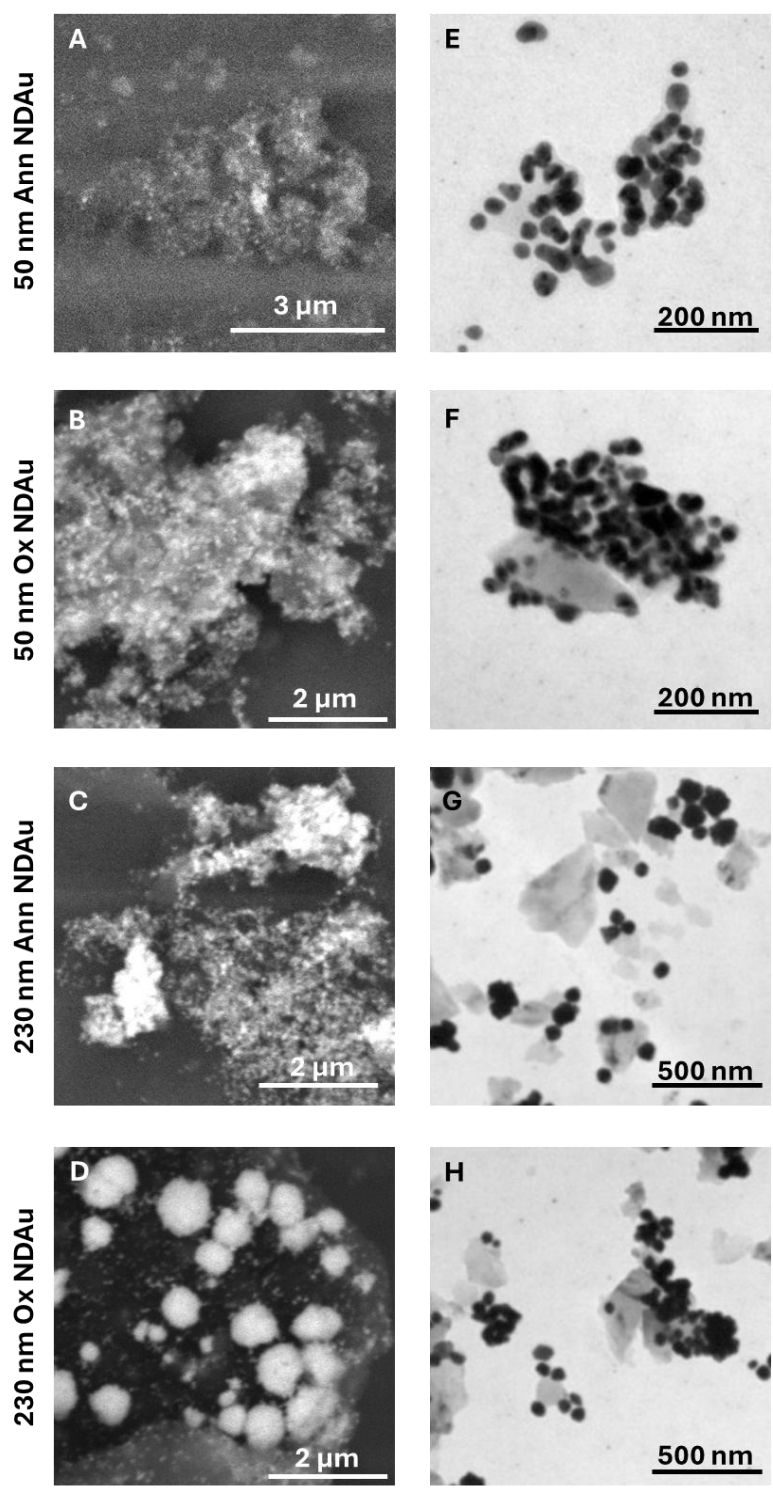

Figure 3.5. SEM (left) and TEM (right) images of NDAu annealed (Ann) and oxidized (Ox). 50 nm Ann NDAu (A,E); 50 nm Ox NDAu (B,F) 230 nm Ann NDAu (C,G); 230 nm Ox NDAu (D,H).

The ζ-potential of nanoparticles reflects the electric potential at the interface between the particle surface and the surrounding liquid medium. This potential arises from the distribution of ions around the nanoparticle, particularly within the electric double layer, and serves as a key indicator of colloidal stability. As shown in Table 4.4, gold coating resulted in a significant decrease in the ζ-potential of all annealed NDAu samples, shifting toward more negative values. This trend was particularly evident for the AnnNDau 230, where the ζ-potential decreased by almost 35 mV (from -3.84 mV to -39.3 mV), compared to a decrease of around 30 mV for the Ann NDAu 50 (from -26.1 mV to -56.9 mV). This change correlated with better dispersibility in suspension, suggesting that surface modification enhanced the stability of the NDAu. Indeed, it is well-known that nanoparticles with high absolute ζ-potential values tend to repel each other more strongly, reducing aggregation and improving stability, while values close to 0 mV (±5 mV) indicate weaker repulsion and potential instability [69]. In contrast, the ζ-potential of Ox NDAu 50 remained relatively but increased for the Ox NDau 230batch. These variations suggest that while surface modification influences ζ-potential, factors such as particle size and surface-to-volume ratio also play a role, with smaller NDAu samples consistently exhibiting more negative ζ-potential values.

**Table 5** Zeta Potential for all the annealed (Ann) and oxidized (Ox) Nanodiamonds (ND).

| Sample | Zeta Potential, ζ (mV) |
|---|---|
| **Ann ND 50** | - 26.1 ± 0.6 |
| **Ox ND 50** | -54.8 ± 1.8 |
| **Ann ND 230** | -3.8 ± 0.2 |
| **Ox ND 230** | -67.7 ± 1.1 |
| **Ann NDAu 50** | - 56.9 ± 1.0 |
| **Ox NDAu 50** | -54.9 ± 0.2 |
| **Ann NDAu 230** | -39.3 ± 0.6 |
| **Ox NDAu 230** | -40.8 ± 0.8 |

The observed changes in the ζ-potential of NDAu samples can be significantly influenced by the presence of polyphenolic compounds and carboxyl groups introduced during the synthesis process. These functional groups, commonly found in plant extracts used for nanoparticle synthesis, play a crucial role in determining the surface charge and stability of nanoparticles. Polyphenols, abundant in many plant extracts, contain phenolic hydroxyl (-OH) and carboxyl (-COOH) groups. At physiological pH, these groups can deprotonate, imparting a negative charge to the nanoparticle surface. This increase in negative surface charge enhances the electrostatic repulsion between particles, thereby improving colloidal stability. For instance, a study on chitosan-functionalized hydroxyapatite nanoparticles demonstrated that polyphenols provide a negative ζ-potential due to the deprotonation of phenolic and carboxylic groups at pH 7.4, leading to increased stability of the colloidal system [70,71]. The surface charge of nanoparticles, as indicated by ζ-potential, is a critical factor in their interaction with cellular membranes. Cell membranes typically possess a net negative charge due to the presence of phospholipid bilayers. Nanoparticles with highly negative ζ-potentials may experience electrostatic repulsion, potentially reducing cellular uptake. Conversely, nanoparticles with positive ζ-potentials can interact more readily with cell membranes, facilitating uptake but also posing a risk of cytotoxicity due to membrane disruption. A study investigating polymeric nanoparticles found that those with positive surface charges exhibited stronger interactions with cells and higher cytotoxicity compared to negatively charged nanoparticles [72- 74].

Different extracts contain varying amounts and types of polyphenols and other functional groups, which can modulate the surface chemistry and charge of the nanoparticles, thereby impacting key aspects of NDAu stability and its interactions with cells. Given the influence of size, ζ-potential, and other physicochemical characteristics of nanoparticles on their stability and cellular interactions, understanding these surface properties is essential for evaluating NDAu's performance in biomedical applications. Accordingly, we investigate the biological evaluation of NDAu, focusing on cell viability and cytotoxicity assays, cell survivability and clonogenic assays, and NDAu cellular uptake. These experiments offer valuable insights into the biological behavior of NDAu and their potential applications in radiosensitization and drug delivery.

*.3.2 Cell Studies*

3.2.1. Cell Viability

To evaluate the cytotoxicity of the newly synthesized NDAu, MTT assay was performed on A549 and PANC-1 cell lines. For the study of the viability, in both cells lines, we used concentrations ranging from 10 to 100 µg/mL.

The results, for both cells after incubation for 24 and 48 hours, indicate that the 50 nm NDAu did not exhibit statistical differences in viability, as shown in Figure 5. This outcome is not unexpected since both gold and diamond are biocompatible materials. The PANC-1 cell line showed a lower viability at 100 µg/mL, for Ann NDAu 48h post-incubation.

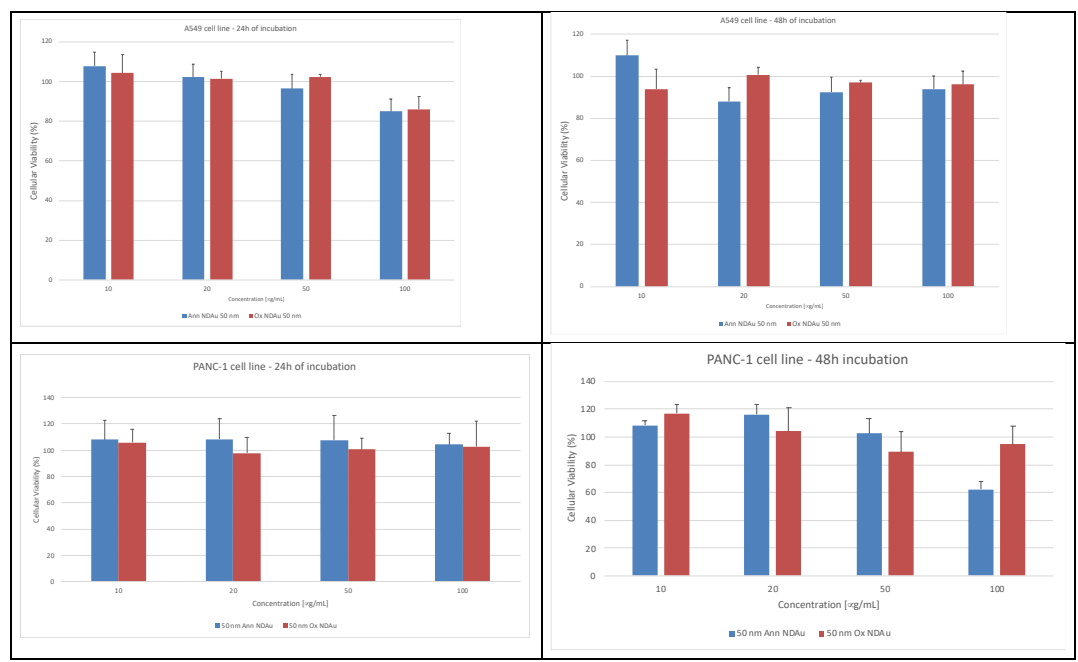

Figure 5 Cellular viability of A549 and PANC-1 cell lines exposed to various concentrations of annealed (Ann) and oxidized (Ox) 50 nm NDAu nanoparticles incubated for 24 hours (left) and 48 hours (right), as measured by the MTT assay. Viability values are normalized to the control, and error bars represent the standard error of the mean (SEM).

In contrast, the NDAu 230 samples presented a slightly cytotoxicity after 48 hours of incubation than their 50 nm counterparts. The viability Ann and Ox NDau 230was on average 70% and 60%, for 24h and 48h, respectively. This decrease in cellular viability may be attributed to their larger size, ability to aggregate and lower dispersibility. These results suggest that 230 nm NDAu may be deposited on the surface of cells and interfere with cell function.

3.2.3. – NDAu cellular uptake

Cellular uptake of NDAu was evaluated using gold as a marker for the nanoparticles. Due to their better biological activity, only the Ann and Ox NDAu 50 uptakewas analyzed. The results for A549 cells incubated with 20 µg/mL of Ann and Ox NDAu 50 are presented in Figure 7.

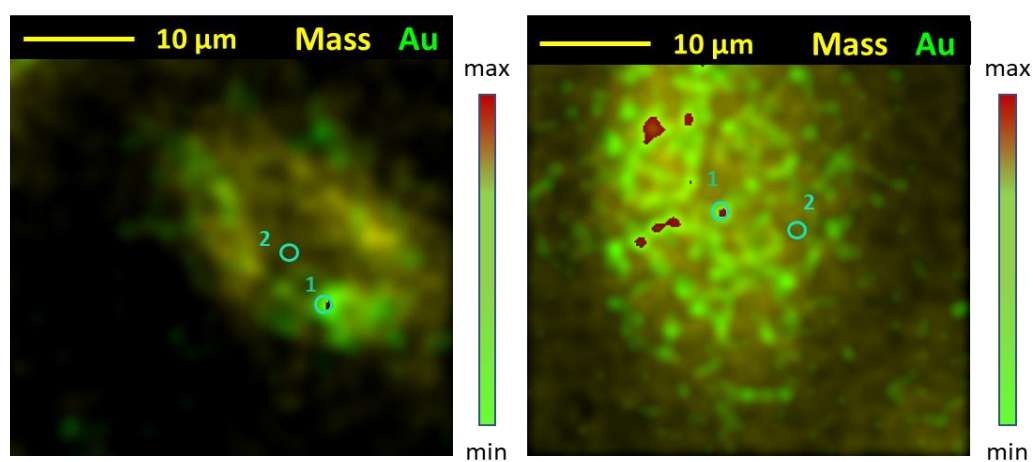

Figure 7. Nuclear microscopy superimposed images of mass density (STIM) and Au distribution (PIXE) in a A549 cell incubated incubated with 20 µg/mL Ann (left) and Ox (right) NDAu 50. The gradient amount of Au is represented by a colour scale: low-green to high-red. The mass distribution is represented by a yellow positive gradient.

As shown in Figure 8, gold was distributed throughout the cell rather than localized in a specific cellular compartment. Notably, the Ox NDAu exhibited a more uniform gold distribution across the cell and a smoother gold gradient. This observation aligns with the superior dispersibility of Ox NDAu 50 in aqueous environments compared to the annealed counterpart. In the annealed NDAu treated cell large Au deposits are scattered throughout the cell. Using PIXE spectra analysis and incorporating matrix composition and thickness estimates from the EBS data, Au concentrations were quantified in gold hotspots and coldspot, regions with high and low Au intensity respectively, represented by points 1 and 2 in the distribution maps of Figure 8. The quantitative results are reported on Table 6.

Table 6 – Au concentration in the Au hotspot and coldspot in A549 cells incubated with 20 µg/mL Ann and Ox NDAu 50 .

| Nanodiamond | [Au] coldspot | [Au] hotspot |

|  | (mg/g (dry weight)) | (mg/g (dry weight)) |
|---|---|---|
| Ann NDAu 50 | 1.2 ± 0.3 | 5.2 ± 0.6 |
| Ox NDAu 50 | 4.4 ± 0.2 | 7.9 ± 0.5 |

As evidenced in Table 6, the Au distribution across the cell is not uniform. Cells incubated with annealed nanoparticles exhibit a roughly four-fold increase in Au concentration between coldspots and hotspots, while those incubated with oxidized nanoparticles show an approximately two-fold increase. Additionally, the Ox NDAu demonstrates greater cellular uptake than the Ann NDAu, likely due to differences in surface modifications. This is evident from the higher internal Au concentration observed in cells incubated with Ox NDAu. Figure 8 shows a 3D representation of the amount of Au in these cells corrected for beam energy loss at different depths obtained from the EBS spectra fits.

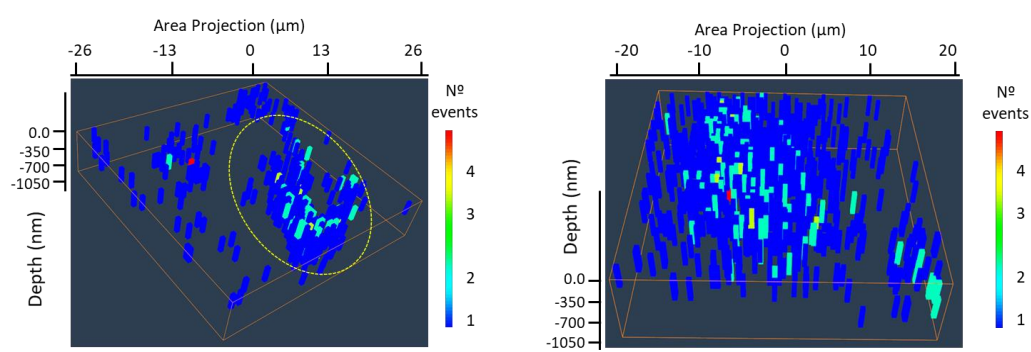

Figure 8. 3D representation of Au distribution in cells incubated with (left) Annand (right) Ox NDAu 50 and represented in Figure 8. A three-layer model of approximately 340 nm each described the distribution of Au through the depth of the cell. The zero depth represents the cell surface. The dotted line in the left image encircles the cell treated with Ann NDAu 50 shown in Figure 8. The amount of Au is expressed as the number of events corrected for beam energy loss at different depths (color scale bar, left). Both images were constructed using the MORIA software. [75]

The 3D representation reveals a higher number of events in deeper cellular layers in Ox NDAu 50 treated cell, confirming gold internalization through the whole cell depth (approximately 900-1000 nm). In the cell treated with annealed NDAu, gold remains in the uppermost surface layer, which is consistent with the Au distribution shown in Figure 8. Although live cells typically have a greater thickness, it is important to note that these analyses are performed in vacuum, which requires the cells to be freeze-dried. Notably, Figure 8 further corroborates the superior dispersion of oxidized NDAu. This may facilitate NDAu uptake, as evidenced by the uniform depth distribution of Au throughout the cell.

The internalization of the particles is further supported by SEM images (Fig. 10), which clearly depict the distribution of nanoparticle agglomerates outside (yellow, Fig. 10 e) and inside (orange, Fig. 10 e) of the A549 cells. Additionally, the cells exhibit the typical morphology of A549 cells under both control conditions and following exposure to Ox NDAu 50 for a short period (24 hours) [76], in alignment with the results regarding the Cell

Viability (Figure 5) . The cell structure remains intact, with normal surface topology and no evident signs of damage. However, after 48 hours of contact with the particles, morphological changes become apparent, suggesting an adverse cellular response to the presence of the nanoparticles.

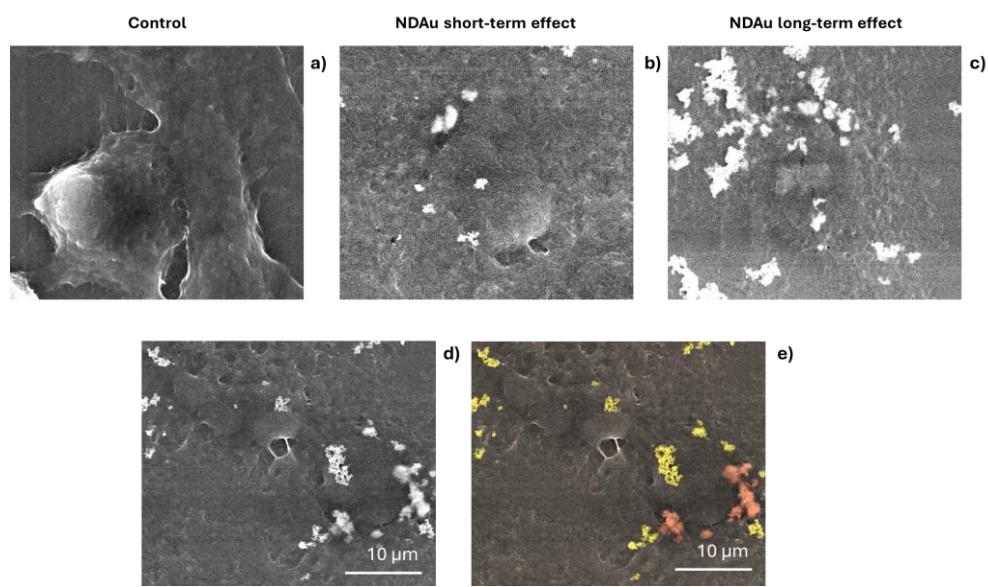

Figure 10 - Scanning Electron Microscopy (SEM) imaging of A549 lung cells: (a) Control cells; (b) cells incubated with 50 μg/mL OxNDAu 50 for 24 hours; (c) cells incubated with 50 μg/mL Ox NDAu 50 for 24 hours followed by 48 hours in fresh medium; (d, e) distribution of Ox NDAu 50 within the cell culture, with yellow indicating Ox NDAu 50 located outside cell structures and orange representing internalized particles.

### 3.2.2. Cell Survival

Figure *6* presents the results of the clonogenic assay performed on the A549 cell line incubated with Ox NDAu 50 for 24 hours. The data indicate that incubation with 20 μg/mL did not interfere with the cells' replicative capacity, as the survival fraction (SF) remained around 90%. However, at a concentration of 50 μg/mL, the survival fraction significantly decreased to approximately 60%. These findings suggest that higher concentrations of Ox NDau 50 may have long-term effects on cellular viability, possibly due to internalization or interactions with the cell membrane. In addition to the survival fraction, colonies formed at 20 μg/mL were larger in size compared to those at 50 μg/mL, further supporting the long-term effects of these nanoparticles on cell replication.

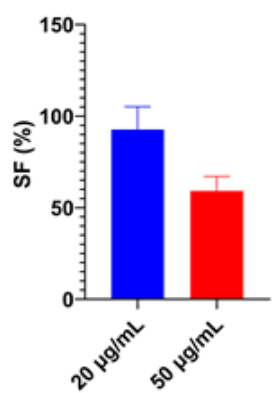

Figure 6 – Survival fraction of A549 cells incubated with 20 or 50 µg/mL oxidized 50 nm NDAu. Results are normalized to the control, and error bars represent the SEM.